\documentclass[
twocolumn, prl,
 showpacs,floats,preprintnumbers,amsmath,amssymb]{revtex4}
\usepackage{epsfig}
\usepackage{amsmath}
\renewcommand{\(}{\left(}
\renewcommand{\)}{\right )}
\renewcommand{\[}{\left [}
\renewcommand{\]}{\right ]}

\def\pa{\partial}

\def\bea{\arraycolsep .1em \begin{eqnarray}}
\def\eea{\end{eqnarray}}

\let\be=\beta
\let\no=\nonumber

\def\eq#1{Eq.(\ref{#1})}
\def\refr#1{\cite{#1}}

\def\eqs#1{Eqs.(\ref{#1})}
\def\s0#1#2{\mbox{\small{$ \frac{#1}{#2} $}}}
\def\0#1#2{\frac{#1}{#2}}

\def\cpl#1#2#3{Chin. \ Phys.\ Lett. \ {\bf #1}, #2 (#3)}
\def\ctp#1#2#3{Commun.\ Theor.\ Phys. \ {\bf #1}, #2 (#3)}

\def\prv#1#2#3{Phys. Rev.  {\bf #1}, #2 (#3)}
\def\pra#1#2#3{Phys. Rev.  {\bf A #1}, #2 (#3)}

\def\prl#1#2#3{Phys. Rev. Lett. {\bf #1}, #2 (#3)}
\def\ann#1#2#3{Ann. Phys. (N.Y.) {\bf #1}, #2 (#3)}

\def\prep#1#2#3{Phys.\ Rep.\ {\bf #1}, #2 (#3)}

\def\rmp#1#2#3{Rev.\ Mod.\ Phys.\ {\bf #1}, #2 (#3)}

\def\sci#1#2#3{Science\ {\bf #1}, #2 (#3)}
\def\nat#1#2#3{Nature\ {\bf #1}, #2 (#3)}
\begin{document}
\title{Three-dimensional correlated-fermion phase separation from
    analysis of the geometric mean of the individual susceptibilities}
\author{Ji-sheng Chen\footnote{chenjs@iopp.ccnu.edu.cn},~~~Fang Qin,~~~~and Yan-ping Wang}
\address{Institute of Particle Physics,
Central China Normal University, Wuhan 430079, People's Republic of
China}
\begin{abstract}
A quasi-Gaussian approximation scheme is formulated to study
    the strongly correlated imbalanced fermions thermodynamics, where the mean-field theory is not applicable.
The non-Gaussian correlation effects are understood to be captured
    by the statistical geometric mean of the individual susceptibilities.
In the three-dimensional unitary fermions ground state, an {\em
    universal} non-linear scaling transformation relates the physical
    chemical potentials with the individual Fermi kinetic energies.
For the partial polarization phase separation to full polarization,
    the calculated critical polarization ratio is $P_C={[1-(1-\xi)^{6/5}]}/{[1+(1-\xi)^{6/5}]}\doteq
    0.34$.
The $\xi=4/9$ defines the ratio of the symmetric ground
    state energy density to that of the ideal fermion gas.

\end{abstract}



\maketitle

Phase separation resulting from the microscopic dynamics involves a
    wide range of natural phenomena. Currently,
it also serves as the pivotal topic in the strongly
    interacting fermions quantum many-body theory.

In the past few years, considerable efforts understanding the
    crossover physics from Bardeen-Cooper-Schrieffer to Bose-Einstein
    condensation(BCS-BEC crossover) with
    ultra-cold atomic fermi gases have been made.
At the Feshbach resonance point,
    the divergent $S$-wave scattering length with the existence of a zero-energy bound state
    can exhibit the thermodynamic
    universality\refr{Ho2004,Giorgini2007}.

Another exciting issue is on the
    asymmetric fermions phase diagram with unequal populations,
    which  persists as a fundamental
    theme for a long time\refr{Sarma,Fulde,Larkin}.
With the Feshbach resonance techniques,
    tuning interaction strength and
    controlling the population or mass imbalance among the components
    offer a playground to test the non-perturbative many-body theory for
    the asymmetric fermions thermodynamics\refr{Partridge2006,Zwierlein}.
The experiments show that at zero temperature the system can undergo
    a first-order quantum phase transition from a fully-paired
    superfluidity to a partially polarized normal gas\refr{Yong2007}.
Furthermore, when the imbalance population ratio
    $P=(N_\uparrow-N_\downarrow)/(N_\uparrow+N_\downarrow)$ between the
    two spin components reaches a critical value $P_C$,
    the state can transfer from the partially polarized phase separation to the fully
    polarized normal one.

The three-dimensional strongly interacting
    asymmetric fermions ground state is one of the prominent unresolved problems\refr{Pethick2002}.
Meanwhile, the zero-temperature ground state energy of the
    partial polarization phase separation is of particular importance for explaining experiments\refr{Yong2007,Yong2008}.
For the equally populated spin states, the Monte Carlo methods are
    believed to be capable of doing the simulation calculations.
For the asymmetric system, the numerical simulations will suffer
from the serious fermion sign problem\refr{Braaten2008}. Making a
definitive theoretical conclusion or finding an unbiased
    analytical solution remains as an intriguing but herculean task.
As far as we are aware, until recently, any analytical
    attempt was not known.

There can exist a relation between the $P_C$ and dimensionless
    constant $\xi$ at unitarity with
    $|a|=\infty$\refr{Cohen2005,reddy,Bulgac2007,Yong2008,He2008}.
The $\xi$ defines the ratio of the symmetric fermions energy
    density to that of the ideal noninteracting ones($T=0$).
By exploring the complex non-Gaussian correlations, the purpose of
this \textit{Letter} is offering a novel tool to calibrate the
    ground state energy of strongly interacting asymmetric fermions and
    fix the relation between $\xi $ and $P_C$.
The method is based on extending
    the formalism developed in\refr{chen2007,chen2008,chen2007-1}.
The calculations are performed with $k_B=\hslash=1$.

To address the imbalanced fermions thermodynamics,
   we formulate the universal medium-scaling Hamiltonian
\bea\label{Hamiltonian}
    \tilde{H}=&&-\int d^3x
    \psi_\alpha ^*(x) (\0{\nabla ^2}{2m_\alpha}-\mu_{r\alpha}[n,T])\psi _\alpha (x)\no\\
    &&+\0{U_{\mbox{eff}}^*[n,T]}2 \int d^3x\psi_\alpha ^*(x)\psi^*_\beta (x)
    \psi_\beta(x)\psi_\alpha (x).
    \eea
In \eq{Hamiltonian},
    $\alpha,\be=\uparrow (a), \downarrow (b)$
    represent the hyperfine spin projection Ising-variables.
The Hamiltonian is the same as the Bethe-Peierls
    contact interaction,
    except that the bare coupling constant
    $U_0={4 \pi a}/{(2 m_r)}$  is substituted by a medium regulated functional $U_{\mbox{eff}}^*[n,T] $.
The $m_r={m_a
        m_b}/(m_a+m_b)$ is the reduced mass.
In the vacuum limit $n\rightarrow 0$,
    the $\tilde{H}$ reduces to the original version possessing a global $U(1)$ or $\mathbb{Z}_2$ gauge
    symmetry (assuming $m_a=m_b$ without a loss of generality).

Due to the medium dependence of $U_{\mbox{eff}}^*$,
    the non-trivial $\delta {\cal H}\propto \mu_{r,\alpha}[n,T]N_\alpha$ enforce the energy-momentum
    conservation law\refr{Brown2002}.
Routinely, the strong correlation effects are further
    understood as a spontaneously generated one-body correlation potential from the density
    functional theory viewpoint\refr{chen2007,chen2008},
    with which the general thermodynamic consistencies are strictly guaranteed within the
    non-perturbative procedures.

The medium-scaling formalism allows a natural implementation of the
    strong correlation physics.
Initially, we need to constitute the
    $U_{\mbox{eff}}^*$ functional.
The main observation is that the environment frustrating
    characteristic can be realized by the twisted contact interaction
\bea\label{action}
    U_{\mbox{eff}}^*&&=\0{U_0}{1-\chi_{ab} U_0}.\eea
With \eq{action}, one can define the in-medium inverse scattering
    length notation $a_{\mbox{eff}}$ according to $U_{\mbox{eff}}^*\equiv4
    \pi a_{\mbox{eff}}/m$.

With the ``negative" sign in the denominator of \eq{action},
    the interaction is alternatively renormalized by the strong correlation
    effects or the background fluctuations induced in the counterpart Fermi sea.
To a great extent, the susceptibility $\chi_{ab}$
    is interpreted as an external field modulator simulating the mutual non-linear
    response. Economically, it characterizes the complications of the single particle
    spectrum properties and correlating couplings.

At unitarity, the scaling thermodynamic quantities
    depend solely on the susceptibilities.
To quantify the complicated correlation effects beyond the Gaussian
    statistics,
    the global variable of interest is
    captured by the statistical \textit{geometric mean}(instead of
    arithmetic mean) of the individual $\chi_a$ and $\chi_b$ for the
     $\uparrow$ and $\downarrow$ subsystems
\bea\label{pol}
    \chi_{ab}=\sqrt{\chi_a\chi_b}.
\eea

The susceptibility $\chi_i$ itself can be separately
    calculated from the Lindhard correlation response function in terms of the random phase
    or spin ``wave" approximation, which is also the final goal\refr{chen2007-1}.
Alternatively, it is more convenient to calculate the susceptibility
    $\chi_{i}$ with the generalized Ward-Identity
\bea\label{chi}
     \chi_{i}=\(\0{\pa n_{i}}{\pa \mu
    ^*_{i}}\)_T=\0{1}{T\lambda^3}f_{\012 }(z'_{i}), ~~~~~i=a,b.
\eea
The $f_{j}(z'_{i})$ is the
    Fermi integrals with $j=-\012,\012,\cdots$ and the thermodynamical de
    Broglie wavelength is $\lambda =\sqrt{{(2\pi)}/{mT}}$.
The collective dynamical variables $\mu^*_i$ are defined by the gap
equations
    \eqs{chemical}.
The {\em effective fugacity} $z_i'=e^{\be\mu^*_i}$ is analogous to
    the fugacity $z=e^{\beta \mu}$\refr{chen2007,chen2008}.

At $T=0$, $\chi_{ab}$ is the density of states(DOS) geometric mean
\bea
    \chi_{ab}=\sqrt{N(\epsilon_a)N(\epsilon_b)}.
\eea Here, $N(\epsilon_i)$ is the familiar un-perturbated DOS near
the Fermi surface for the one component fermions\refr{Pethick2002}.


From the action \eq{Hamiltonian},
    the grand thermodynamical potential $\Omega (T,\mu_a,\mu_b)$ or pressure can be
    presented as the coupled parametric equations with the
   instantaneous quasi-Gaussian approximation method\refr{chen2007,chen2008}
\bea\label{pressure}
    P&&= \0{T}{\lambda^3}\sum _{i=a,b}f_{5/2 }(z'_{i})+\0{4\pi
    a_{\mbox{eff}}}{m}n_an_b+\sum_{i=a,b}\mu_{r,i} n_i,\\
    \label{chemical}\mu_{a}&&=\mu^*_{a}+\0{4\pi
        a_{\mbox{eff}}}{m}n_{b}+\mu_{r,a},~~~~
    \mu_{b}=\mu_a({a\rightleftharpoons b}).
\eea

The dynamical gap equations \eqs{chemical} for the
    corresponding single particle Green function give the definition of
    the effective chemical potential $\mu^*_i$, respectively.
The number density
    $n_i$ is expressed in terms of the defined quasi-particle Fermi-Dirac
    distribution function $f_{k,i}$
\bea
    n_i&&=\0{1}{\lambda^3 }f_{\032 }(z'_{i}),~~~~
    f_{k,i}=\01{z'^{-1}_ie^{\be \0{{\bf
    k}^2}{2m}}+1}.
 \eea

Sticking to the effective interaction \eq{action} renormalized by
    the mixing susceptibility \eq{pol} with \eq{chi},
    the strengths of the correction terms to the Gaussian term are self-consistently
    derived\refr{chen2007,chen2008}
\bea \mu_{r,a}&&={\cal C} (T,z'_a)
    \(\0{4\pi
    a_{\mbox{eff}}}{m}\)^2\sqrt{\0{\chi_b}{\chi_a}}n_an_b, \no\\
    \mu_{r,b}&&=\mu_{r,a}(a\rightleftharpoons b),~~~~
        {\cal C} (T,z'_i)=\0{f_{-1/2}[z'_i]}{2 Tf_{1/2}[z'_i]}.\no
\eea

From the underlying grand thermodynamical potential \eq{pressure}
    and with \eqs{chemical},
    one can derive the remaining thermodynamical
quantities. For example, the energy density $\epsilon=E/V$ and
entropy density $s=S/V$ read
\bea\label{energy-den}
    \epsilon
    &&=\0{3T}{2\lambda ^3}\sum _{i=a,b} f_{5/2}(z'_i)+\0{4\pi
a_{\mbox{eff}}}{m}n_an_b+ T s_r,\\
\label{entropy-den} s&&=\01{{2 T \lambda ^3}}\sum _{i=a,b}\(5 T
f_{5/2}(z'_i)-2 \mu^*_if_{3/2}(z_i')\)+s_r,\no
\eea with
\bea
&&s_r=4n_an_b \(\0{2\pi a_{\mbox{eff}}}{m}\)^2
    \[\sqrt{\0{\chi _b}{\chi _a}}{\cal D}(T,\mu^*_a)+(a\rightleftharpoons b)\],\no\\
    &&{\cal D}(T,\mu^*_i)=\0{\(\0{\pa ^2n_i}{\pa {\mu^*_i}^2}\)_T\(\0{\pa
    n_i}{\pa T}\)_{\mu ^*_i}-\0{\pa ^2n_i}{\pa T\pa {\mu ^*_i}}\(\0{\pa n_i}{\pa
    {\mu^*_i}}\)_T}{2\(\0{\pa n_i}{\pa \mu^*_i}\)_T}.\no
\eea

A few remarks are re-emphasized on the grand thermodynamical
    potential because it plays an important role.
The $\Omega (T,\mu_a,\mu_b)$ is not the
    naive polynomial expanded according to the bare vacuum
    interaction strength $U_0$.
The collective dynamical variable $\mu ^*_i$ mixes
    the low and high order contributions.
Hidden in the factors ${\cal C} (T,\mu^*_i)$ and ${\cal
    D} (T,\mu^*_i)$ as well as in $\chi_{ab}$,
    the collective correlations are combined with the infinite individual
    dynamical high order effects.
The equations \eq{pressure}-\eq{chemical} are
    highly non-linear although they appear a set of coupled algebra ones.
The $\mu ^*_i$ dependence of $\Omega$
    can be numerically eliminated in favor of the physical chemical potential $\mu_i$.

In the following, we will examine the zero-temperature asymmetric
interacting system thermodynamics, which can be derived
analytically. At $T=0$, the pressure \eq{pressure} and energy
    density \eq{energy-den} expressions read
\bea P&&=4 \(\0{2\pi a_{eff}}{m}\)^2 n_an_b \(C(\epsilon _a)
    n_a\sqrt{\0{k_b}{k_a}}+(a\rightleftharpoons b)\)\no\\
    &&~~~~+\025\sum _{i=a,b} n_i \epsilon
    _i+\0{4\pi a_{eff}}{m}n_a n_b,\\
    \epsilon && =\035\sum _{i=a,b} n_i \epsilon _i+\0{4\pi
    a_{eff}}{m}n_a n_b,
 \eea
where the ``Fermi" kinetic energy $\epsilon _i$,
    particle number density
    $n_i$ and rearrangement factor $C(\epsilon _i)$ are
\bea
    \epsilon _i=\0{k_i^2}{2 m},~~~~~~~n_i=\0{k_i^3}{6 \pi ^2},
    ~~~~~~~C(\epsilon _i) =\01{4\epsilon _i}.
\eea

The reduced chemical potentials \eq{chemical} and $a_{\mbox{eff}}$
are
    \bea \mu_a&&=\epsilon_a+\0{4\pi
    a_{\mbox{eff}}}{m}n_b+C(\epsilon _a)\(\0{4\pi a_{eff}}{m}\)^2
    \sqrt{\0{k_b}{k_a}}n_an_b;\no\\
    \mu_b&&=\mu_a(a\rightleftharpoons b),~~~~
    a_{\mbox{eff}}=\0{a}{1-
    \0{2 a}{\pi}\sqrt{k_a k_b}}.\no
\eea The $k_i$ is the corresponding Fermi momentum.

With unitary $a=|\infty|$,
    the above expressions can be further reduced to an oversimplified compact formalism
\bea
    \label{pressure-1} P=\025 n_a
    \mu_a+ \025 n_b \mu_b,~~\epsilon=\035 n_a \mu_a+ \035
    n_b \mu_b,
\eea
 with
 \bea \label{trans-1}
    \mu_a=\epsilon_a-\059 \epsilon _b \(\0{\epsilon_b}{\epsilon_
    a}\)^{1/4},
~
   \mu_b=\epsilon_b-\059 \epsilon _a
    \(\0{\epsilon_a}{\epsilon_ b}\)^{1/4}.
\eea

The salient feature of the main results \eq{pressure-1}
    is that they appear as the $\uparrow (a)$ component contributions ``plus" those of
    $\downarrow(b)$.
The zero-temperature unitary fermions thermodynamics
    obeys the new form of universality, i.e., the Dalton partial pressure law of the non-interacting
    {\em ideal} gas.
Implicitly, the interaction and collective
    correlation information is incorporated through the chemical
    potentials \eq{trans-1}.

Let us further discuss the physical chemical potentials.
They are
    related with the Fermi kinetic energies of the two subsystems
    through the non-linear transformations \eq{trans-1}.
With \eq{trans-1},
    the energy density and pressure can be expressed as the regular functions of the Fermi kinetic
    energies according to \eq{pressure-1}.

\begin{figure}[ht]
        \centering
        \psfig{file=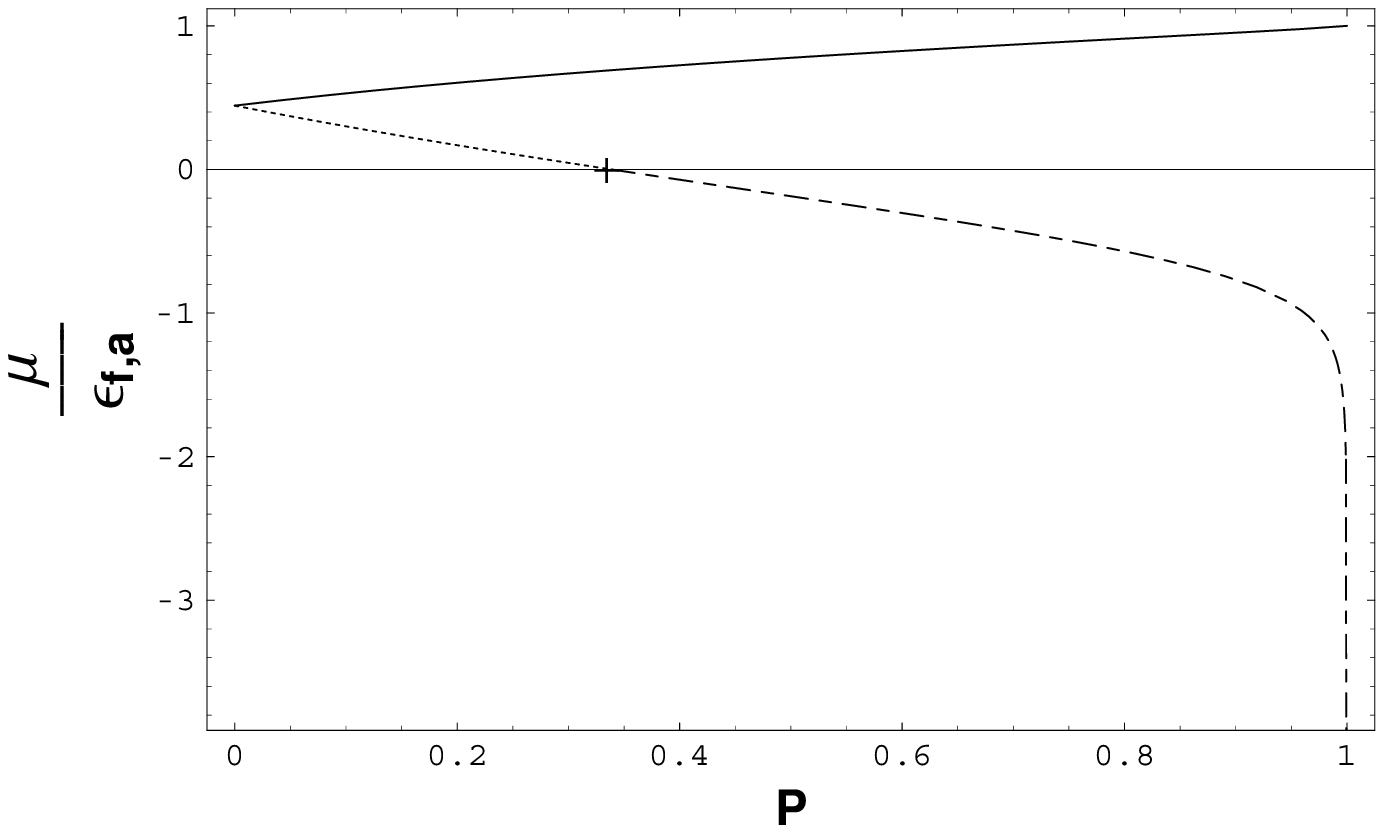,width=6.5cm,angle=-0}
\caption{\small
    Rescaled chemical potentials versus $P$ at unitarity, with ``+"
    indicating the critical position. The above curve is for the
    majority fermions chemical potential while the below one is for the minority
    component.}\label{fig1}
\end{figure}

With the population imbalance polarization ratio $P$ ($0\leq
    P\leq1$, assuming $n_a\geq n_b$ and $\mu_a\geq \mu_b$)
\bea
    P=\0{n_a-n_b}{n_a+n_b},
\eea the numerical solutions of the
    chemical potentials are presented
    in Fig.\ref{fig1}.
One can see the minority chemical potential($\mu_b$)
    decreases very rapidly while crossing the transverse axis.

For $D=3$-dimensional, the reciprocal relations of the
    transformations \eq{trans-1} are very involved; i.e.,
    it is a hard task to express the Fermi kinetic energies in
    terms of the chemical potentials $\mu_a$ and $\mu_b$.
In principle, the Fermi kinetic energies and particle number
    densities in terms of the physical chemical potentials can be the
    functional formalisms such as $\epsilon _a
    [\mu_a,\mu_b, (\0{\mu_a}{\mu_b})^{1/4}]$.
In addition to the mathematical complication, the inverse
    transformation can be singular and nontrivial when $\mu_b=0$.
The minority fermions chemical potential
    $\mu_b$ will change sign at this point indicated by $P_C$.
In other words, the minority component will quickly ``collapse"
    due to the strongly attractive interaction and the system consists
    of a single Fermi surface;
    below it, there can be two distinct Fermi surfaces  and the
    phase separation is favored\refr{Sheehy2007}.

Namely, the vanishing $\mu_b$ determines the
    transferring criterion position from the phase separation-partial
    polarization to the full polarization state.
The analytical relation between $P_C$ and
    $\xi$ can be derived from \eq{trans-1}
\bea
    P_C=\0{1-(1-\xi)^{6/5}}{1+(1-\xi)^{6/5}}.
\eea
With $\xi=\049$,
    the calculated critical ratio is $P_C\doteq 0.34$.
It is in agreement with the Monte Carlo(MC) result $P_C\approx0.39$
    \refr{Lobo2006,Pilati2008}.
The mean-field theory gives $P_C\approx
    0.93$\refr{Sheehy2007}, while
the recent experimental value is $P_C\approx 0.36$\refr{Yong2007}.

\begin{figure}[ht]
        \centering
        \psfig{file=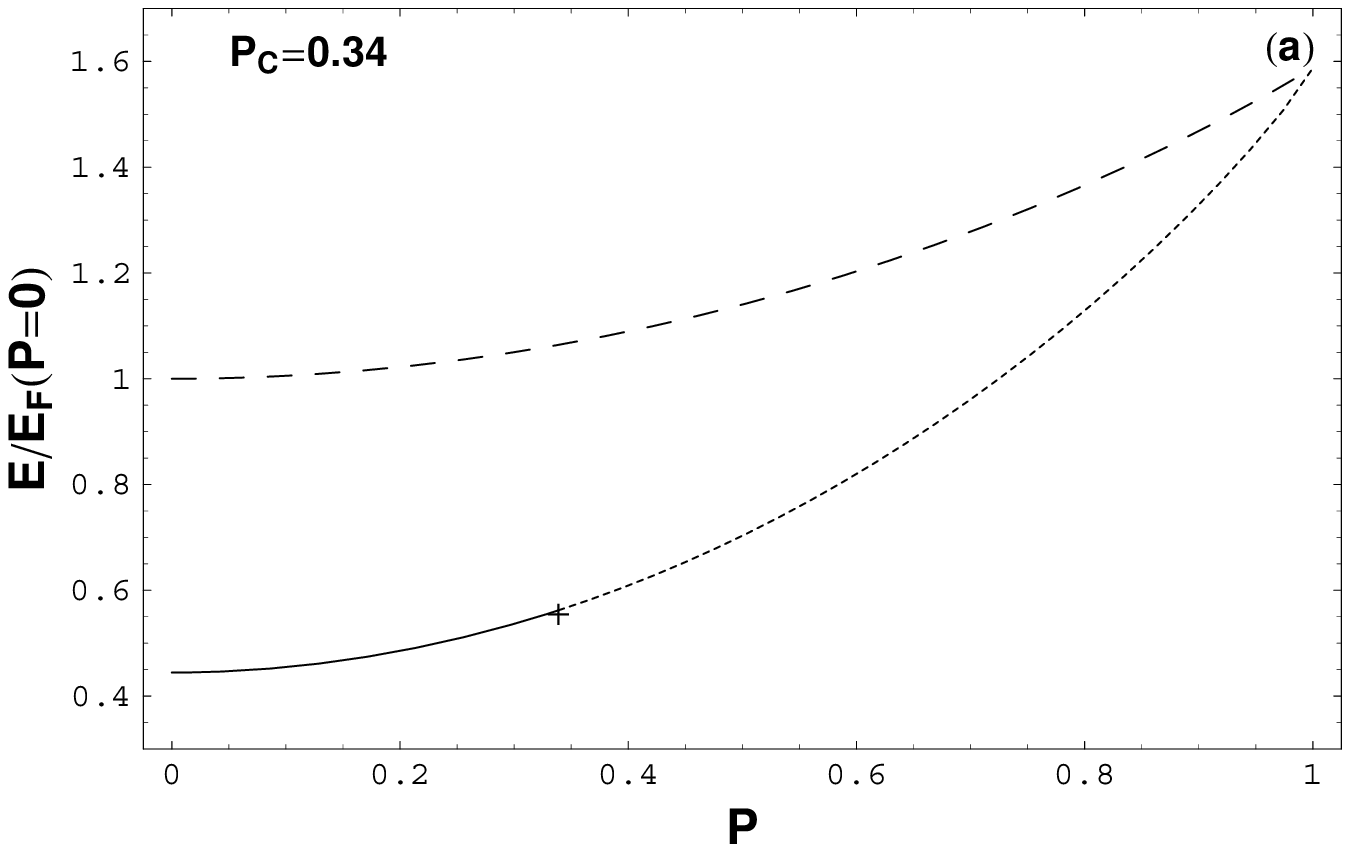,width=6.4cm,angle=-0}
        \psfig{file=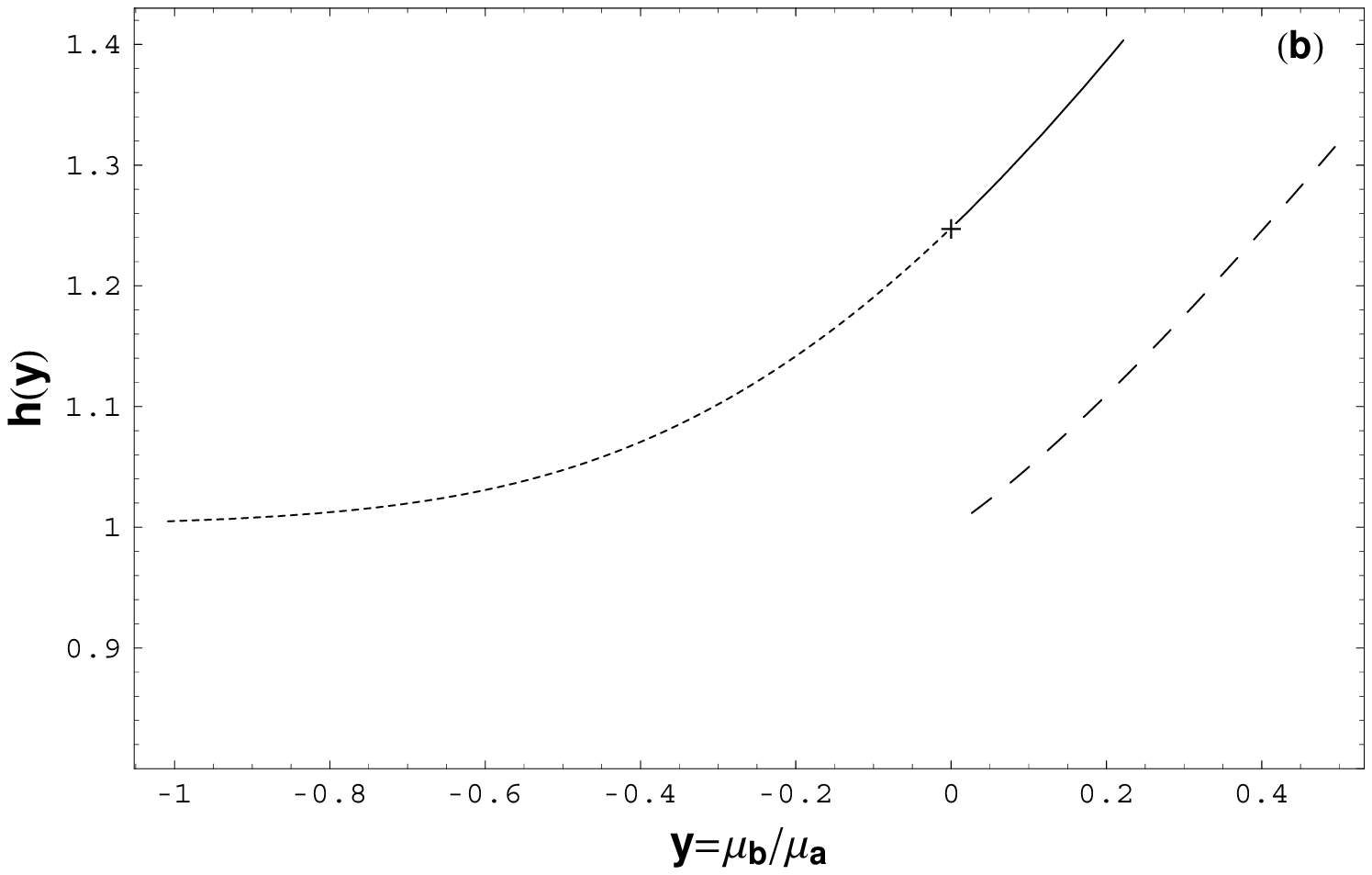,width=6.32cm,angle=-0}
 \caption{
     \small
    a), Ground state energy versus $P$ with fixed density\refr{reddy}.
    b), function $h(y)$ versus $y=\mu _b/\mu_a$\refr{Bulgac2007}. The dashed curve is for the non-interacting ideal Fermi gas. }\label{fig2}
\end{figure}

\begin{figure}[ht]
        \centering
        \psfig{file=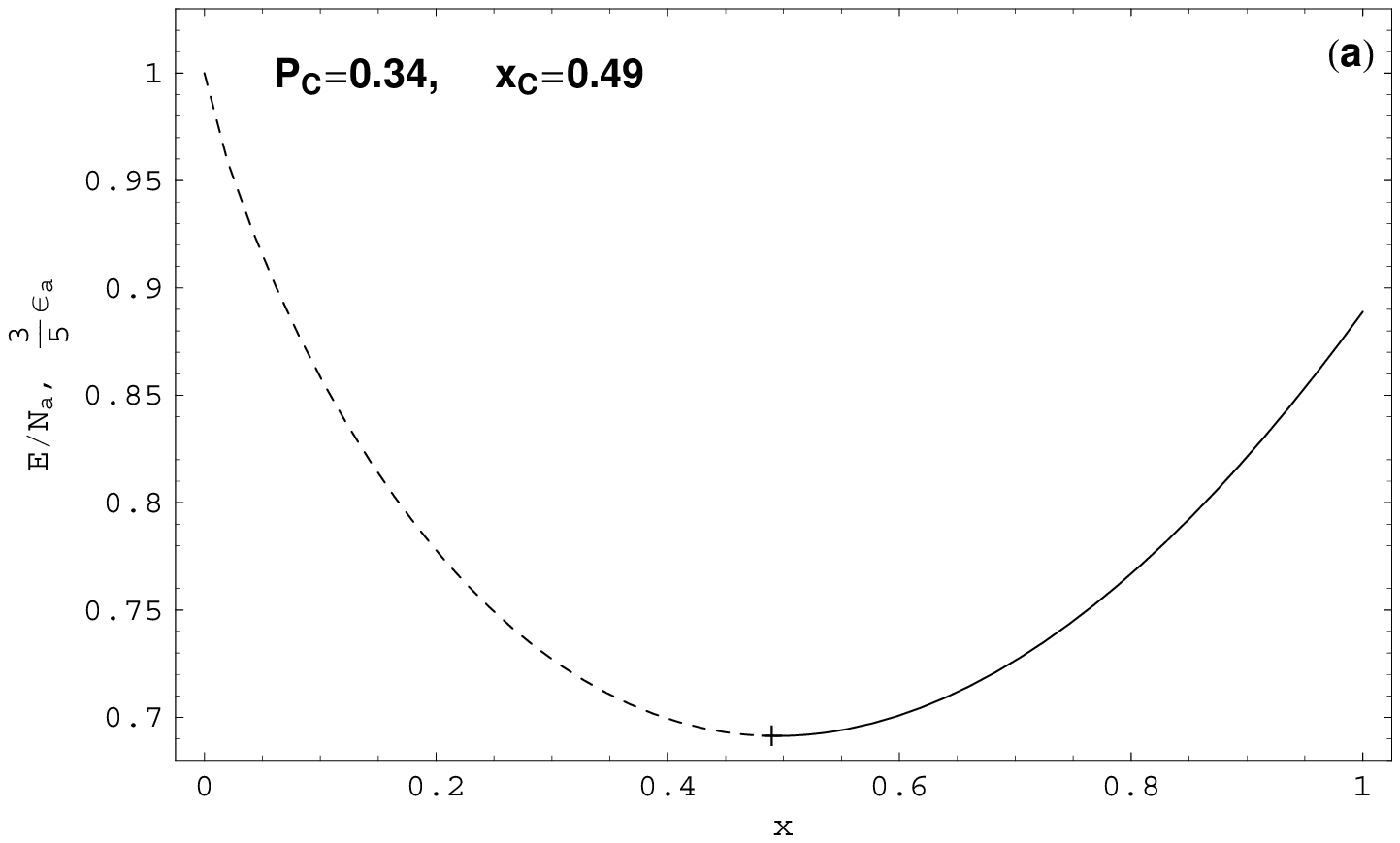,width=6.4cm,angle=-0}
        \psfig{file=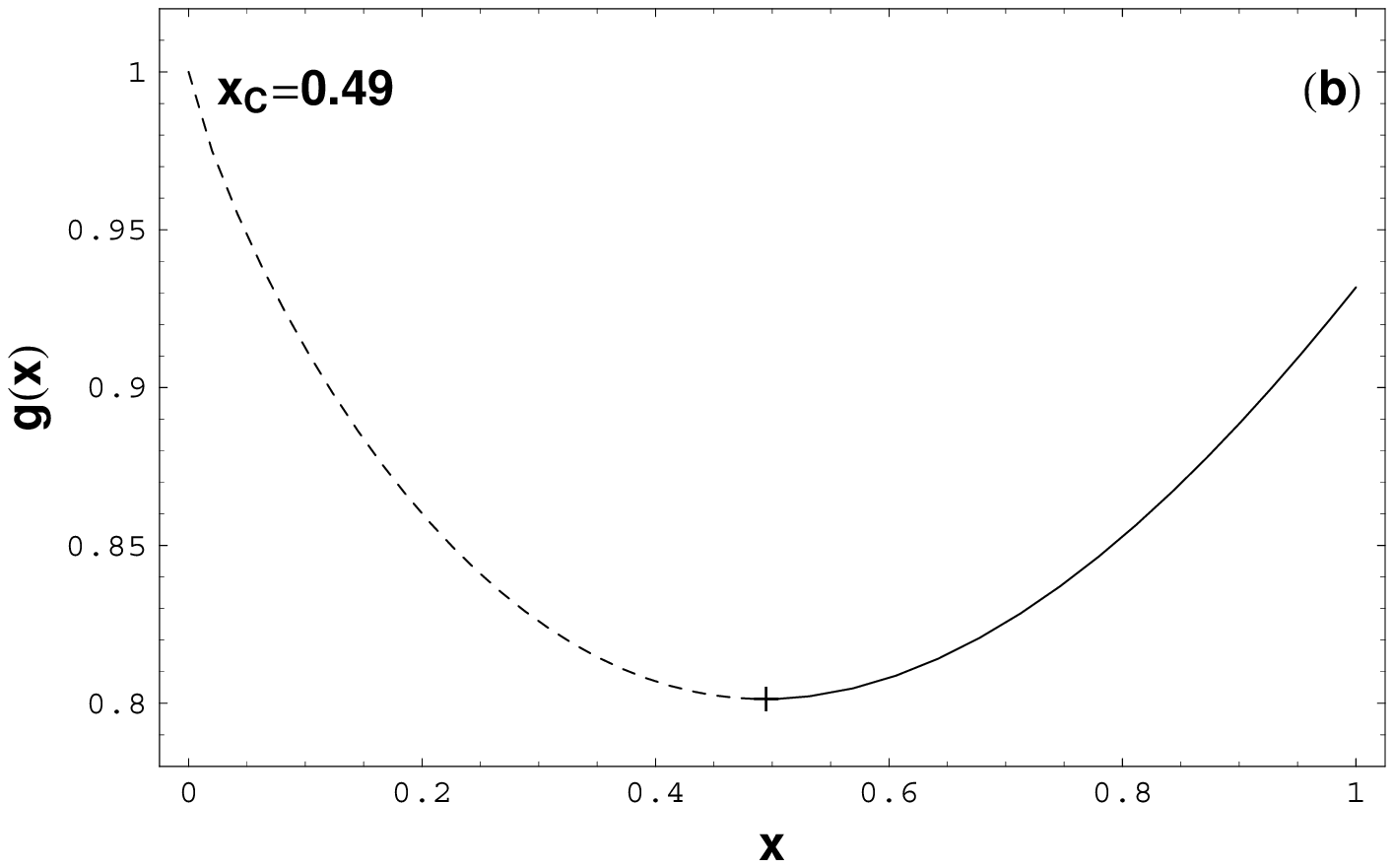,width=6.32cm,angle=-0}
\caption{\small a), $\epsilon/(\035 n_a \epsilon _a)$ versus
$x$\refr{Lobo2006}; b), function $g(x)$  versus
    $x$\refr{Yong2008}.
}\label{fig3}
\end{figure}

The complete ground state energy versus the polarization ratio $P$
    is indicated in Fig. \ref{fig2} and Fig.\ref{fig3}.
They are in agreement with MC calculations very
well\refr{reddy,Lobo2006}.

For the unitary fermions ground state, the equation of state can be
scaled as\refr{Cohen2005,Yong2008,Bulgac2007} \bea \epsilon
    (n_a,n_b)&&=\0{3(6 \pi ^2
    )^{2/3}}{10 m} \[n_ag\(\0{n_b}{n_a}\)\]^{5/3},\\
    P(\mu_a,\mu_b)&&=\0{(2m )^{3/2}}{15\pi ^2} \[\mu_a h\(\0{\mu_b}{\mu_a}\)\]^{5/2}\eea
with the analytical scaling function \bea
    g(x)&&=(1-\0{10}9 x^{5/6}+x^{5/3})^{3/5}. \eea

It is easy to verify that $P_C$ corresponds to the critical
    concentration defined in Refs.\refr{Yong2008,Lobo2006} with $x_C\doteq 0.49$, where the scaling $g(x)$ function takes the minimum value
    as indicated in Fig. \ref{fig3}(b).
The convex behavior of  $g(x)$ coincides with the experimental
     measurement\refr{Yong2008}.
The numerical solution of the $h(y)$ versus $y=\mu_b/\mu_a$ is
    displayed in Fig.\ref{fig2}.b.
These results contribute to understanding the realistic trapped
    system thermodynamics.

To conclude, the complicated non-Gaussian fluctuation and
    correlation effects beyond the canonical Gaussian techniques are encoded
    with the statistical \textit{geometric mean} of the individual susceptibilities.
The simple medium-renormalized  effective action constitutes
    the bridge towards fixing the energy density functional of the strong interaction fermions system.

The non-linear scaling transformation identity \eq{trans-1}
    relates the physical chemical
    potentials with the Fermi kinetic energies.
The reassuring association between the critical proportional
    ratio $P_C$ and universal coefficient $\xi$ agrees well with the
    Monte Carlo calculations  and experimental measurements.
This non-Gaussian correlation perspective will stimulate further
    insights in understanding the novel quantum phase separation dynamics.
\acknowledgments{J.-s Chen acknowledges the discussions with
    L.-y He, D.-f Hou, J.-r Li and X.-j Xia.
   Supported by the Fund of
    Central China Normal University, Natural Science Foundation of China under Grant No. 10875050,
10675052 and MOE of China
    under projects No.IRT0624.}

\end{document}